\documentclass[a4paper,fleqn,usenatbib]{mnras}

\usepackage[T1]{fontenc}
\usepackage{ae,aecompl}

\usepackage{graphicx}	
\usepackage{amsmath}	
\usepackage{amssymb}	
\usepackage{epsfig}
\usepackage[varg]{txfonts}
\usepackage{natbib,twoopt}
\usepackage{rotating}
\usepackage{pdflscape}
\usepackage{epstopdf}
\usepackage{ctable}
\usepackage{romannum}
\usepackage{array}
\usepackage{multirow}
\usepackage{appendix}
\usepackage{booktabs}

\newcommand{\intl}{\textit{INTEGRAL}}

\newcommand{\xmm}{\textit{XMM-Newton}}
\newcommand{\chandra}{\textit{Chandra}}

\newcommand{\sw}{\textit{Swift}}
\newcommand{\ha}{H{\small $\alpha$}}

\newcommand{\hb}{H{\small $\beta$}}

\def\nh{$N_{\rm H}$}

\def\ergs{erg~s$^{-1}$}
\def\ergcms{erg~cm$^{-2}$~s$^{-1}$}
\def\kms{km~s$^{-1}$}
\def\acm2{atoms~cm$^{-2}$}
\def\wm2{W~m$^{-2}$}
\def\mic{$\mu$m}
\def\cm2{cm$^{-2}$}
\def\se1{s$^{-1}$}
\def\nhe{N_{\rm H}}

\def\Ave{A_{\rm V}}

\def\381{IGR~J18381$-$0924}
\def\9173{IGR~J19173$+$0747}
\def\164{IGR~J17164$-$3803}
\def\088{IGR~J18088$-$2741}

\title[Optical and X-ray spectroscopy of \intl\ sources]{Identifying four \intl\ sources in the Galactic Plane via VLT/optical and \xmm/X-ray spectroscopy\thanks{Based on observations performed with European Southern Observatory (ESO) Telescopes at the Paranal Observatory under programme ID~095.D-0972(A)}}

\author[F. Rahoui et al.]{Farid Rahoui,$^{1, 2}$\thanks{E-mail: farid@rahoui.eu (FR)}
John. A. Tomsick,$^{3}$
and Roman Krivonos$^{4}$
\\
$^{1}$European Southern Observatory, Karl-Schwarzschild-Str. 2, 
D-85748 Garching bei M\"unchen, Germany\\
$^{2}$Harvard University, Department of Astronomy, 60 Garden street, 
Cambridge, MA 02138, USA\\
$^{3}$Space Sciences Laboratory, 7 Gauss Way, University of California, 
Berkeley, CA, 94720-7450, USA\\
$^{4}$Space Research Institute, Russian Academy of Sciences, 
Profsoyuznaya 84/32, 117997 Moscow, Russia\\
}

\date{Accepted XXX. Received YYY; in original form ZZZ}

\pubyear{2016}

\begin{document}
\label{firstpage}
\pagerange{\pageref{firstpage}--\pageref{lastpage}}
\maketitle

\begin{abstract}
We report on FORS2 spectroscopy aiming at the identification of four Galactic Plane sources discovered by \intl, \088, \381, \164, and \9173, complemented by \xmm\ spectroscopy for \381. The presence of broad \ion{H}{i} and \ion{He}{i} emission lines and a flat Balmer decrement \ha/\hb\ show that \088\ is a cataclysmic variable located beyond 8~kpc. For \381, the detection of red-shifted \ha\ and \ion{O}{i} emission signatures and the absence of narrow forbidden emission lines point towards a low-luminosity Seyfert 1.9 nature at $z=0.031\pm0.002$. Its \xmm\ spectrum, best-fit by an absorbed $\Gamma=1.19\pm0.07$ power law combined with a $z=0.026_{-0.008}^{+0.016}$ red-shifted iron emission feature, is in agreement with this classification. The likely \164\ optical counterpart is an M2III star at 3 to 4~kpc which, based on the X-ray spectrum of the source, is the companion of a white dwarf in an X-ray faint symbiotic system. Finally, we challenge the accepted identification of \9173\ as an high mass X-ray binary. Indeed, the USNO optical counterpart is actually a blend of two objects located at the most likely 3~kpc distance, both lying within the error circle of the \sw\ position. The first is a cataclysmic variable, which we argue is the real nature of \9173. However, we cannot rule out the second one that we identify as an F3V star that, if associated to \9173, likely belongs to a quiescent X-ray binary. 
\end{abstract}

\begin{keywords}
X-rays: binaries $-$ Stars: novae, cataclysmic variables $-$ Stars: individuals: \088\, \9173, and \164\ $-$ Galaxies: active $-$ Galaxies: individual: \381\ $-$ Techniques: spectroscopic
\end{keywords}



\section{Introduction}

Since its launch in 2002, the \textit{INTErnational Gamma-Ray Astrophysics Laboratory} satellite \citep[\intl,][]{2003Winkler}, has extensively monitored the hard X-ray sky with the Imager on-Board the \intl\ Satellite \citep[IBIS,][]{2003Ubertini}. In the process, more than 600 \intl\ Gamma-Ray (IGR) sources were detected in the 20--100~keV energy range but many are still unidentified \citep[see, e.g.,][]{2012Krivonos, 2016Bird}. Indeed, the large \intl\ positional uncertainty, typically 1--4\arcmin, makes the subsequent use of soft X-ray facilities such as the \textit{Chandra X-ray Observatory} \citep{2002Weisskopf} the \sw\ satellite \citep{2004Gehrels}, or the \textit{X-ray Multi-Mirror-Newton} telescope \citep{2001Jansen} compulsory to detect the soft X-ray counterparts, without the guarantee of unicity. Even in this case, the soft X-ray source may still be associated with several optical or near-infrared counterparts, some of them very faint, which precludes any definitive identification. 
\begin{figure*}
\begin{center}
\begin{tabular}{cc}
\088&\9173\\
\includegraphics[width=8cm,height=7.35cm]{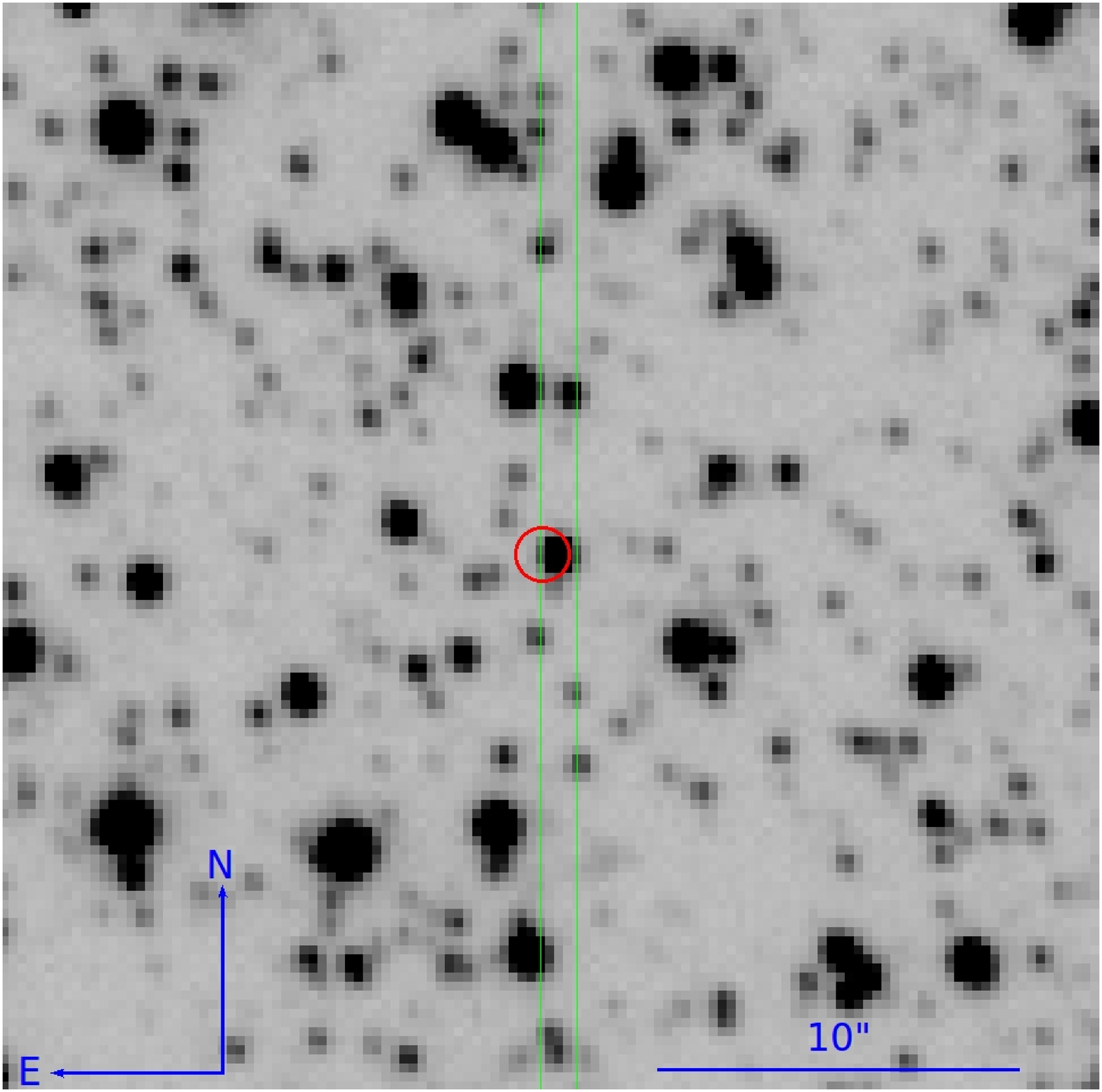}&\includegraphics[width=8cm]{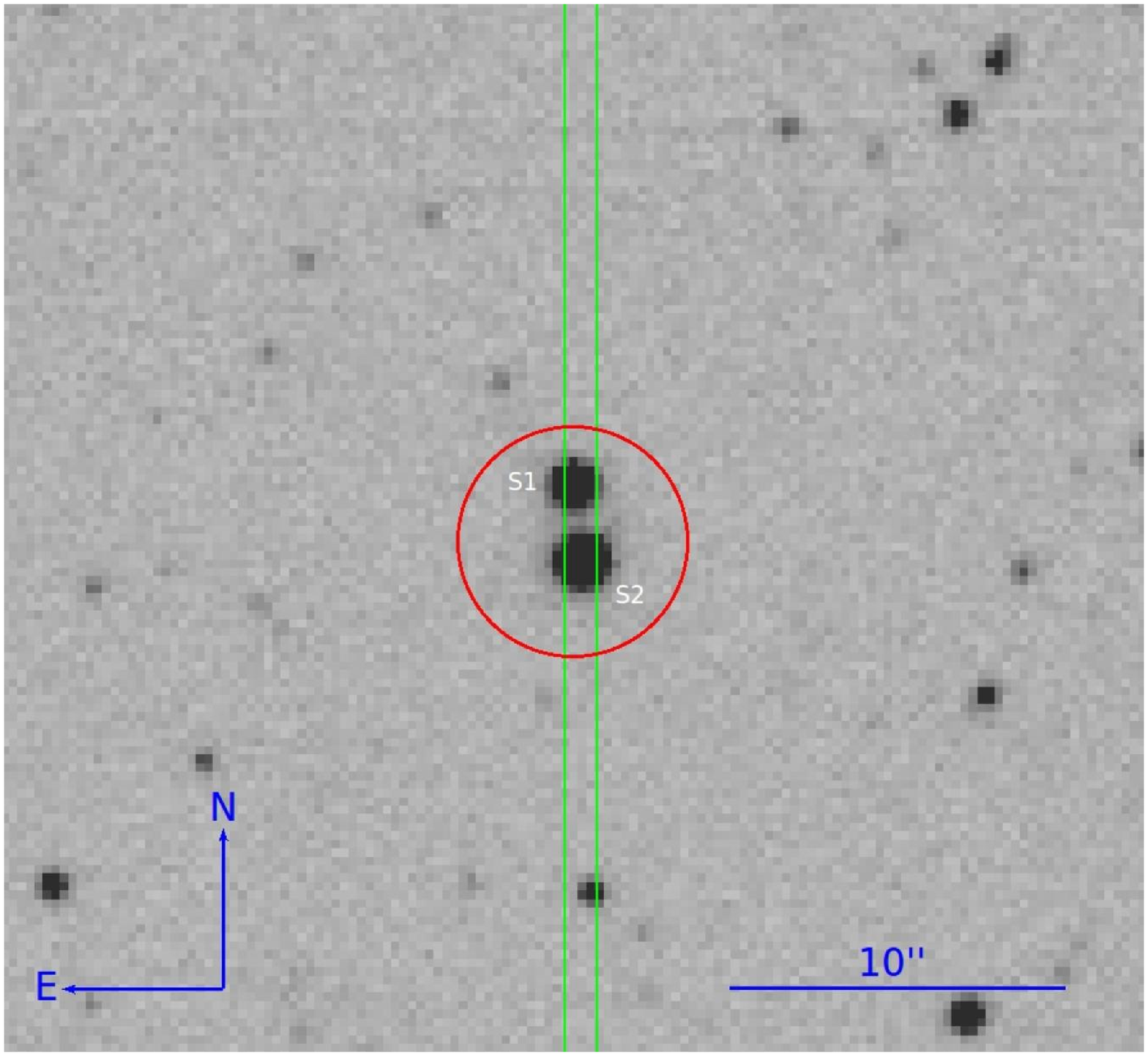}\\
\381&\164\\
\includegraphics[width=8cm,height=7.61cm]{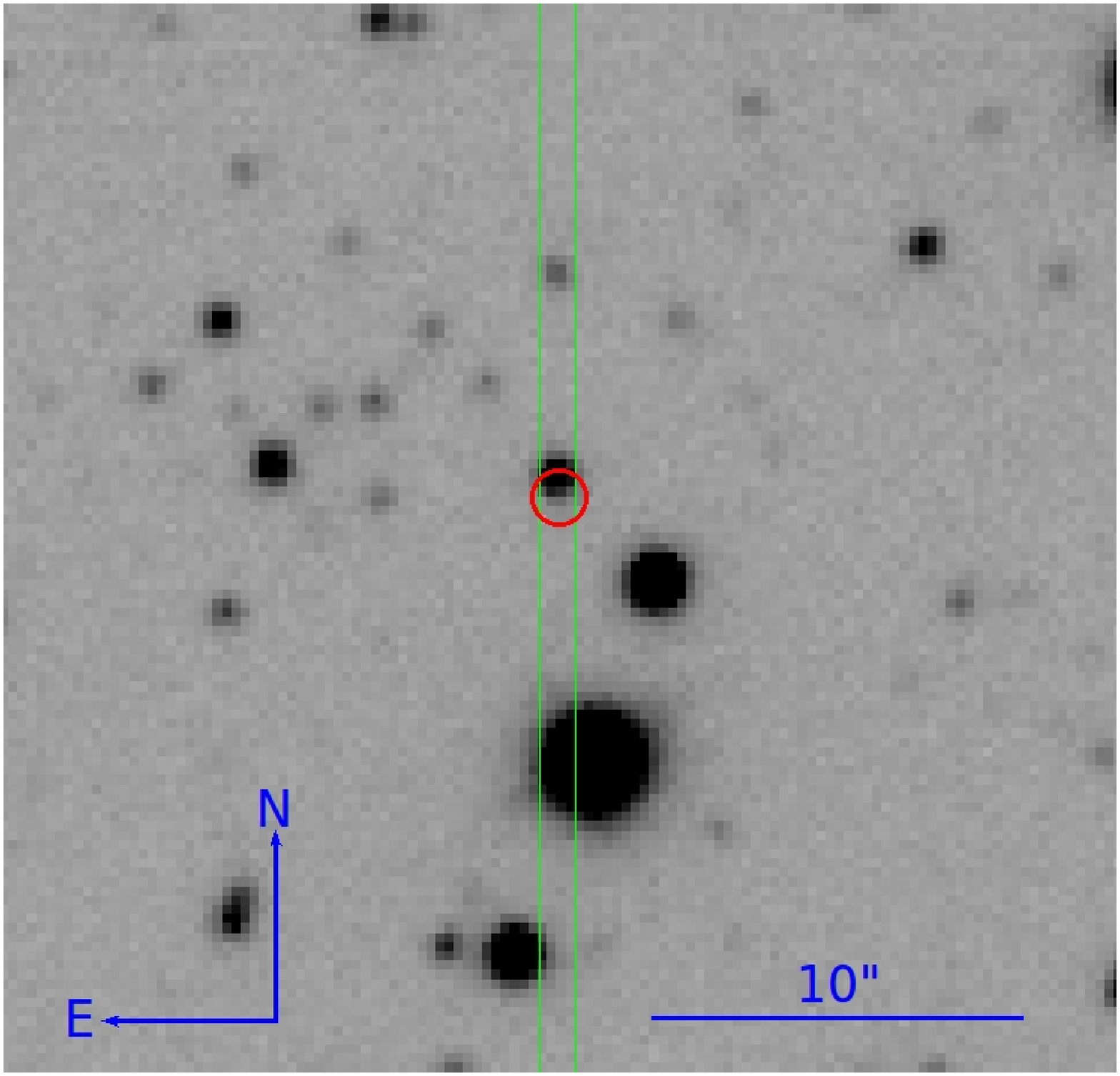}&\includegraphics[width=8cm]{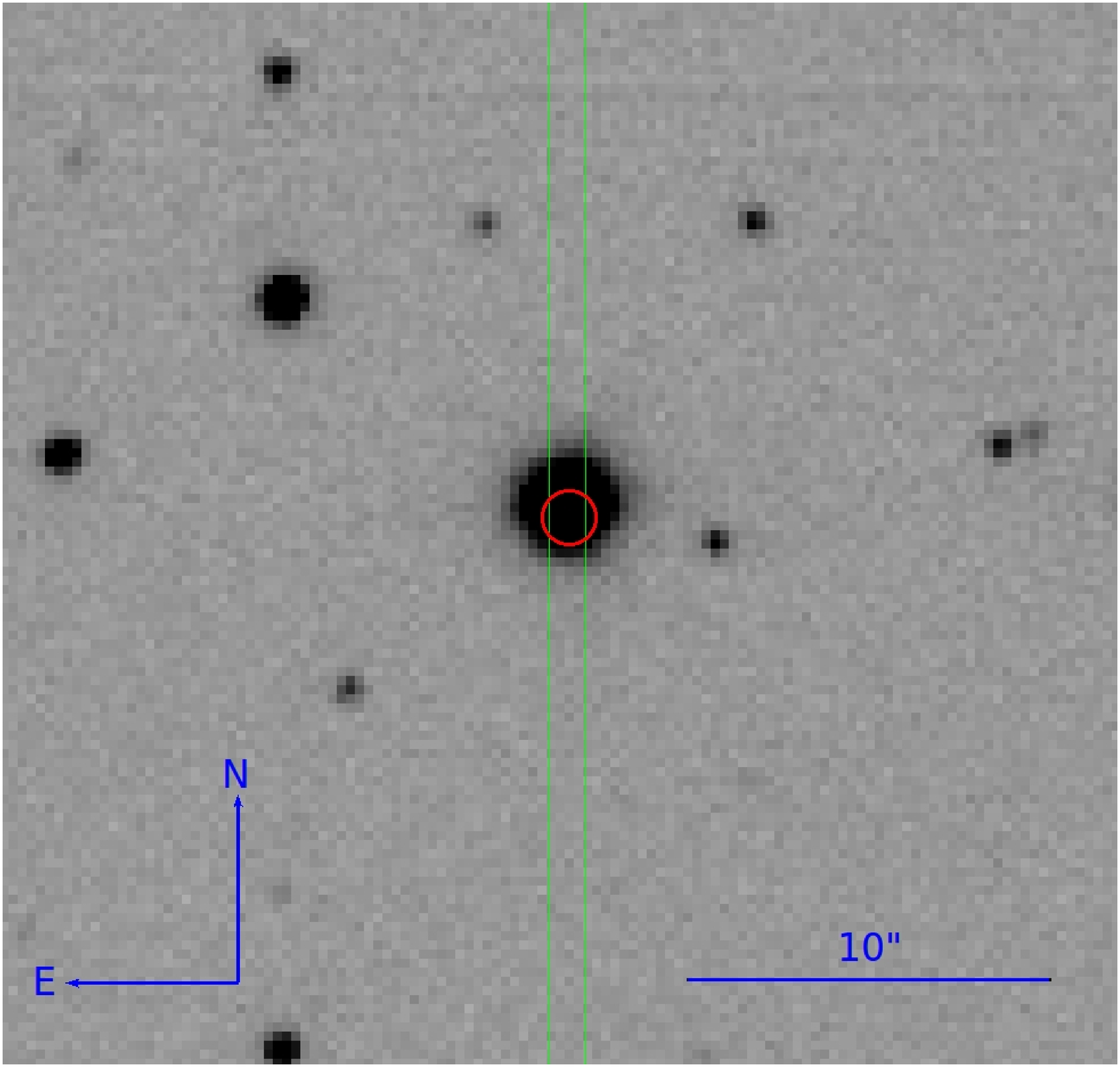}\\
\end{tabular}
\end{center}
\caption{\small FORS2 $V$ acquisition images of the four sources in our sample. \chandra/ACIS error circles (\sw/XRT in \9173\ case) and the 1\arcsec\ slit are superimposed.}
\label{imsources}
\end{figure*}

\begin{figure*}
\begin{center}
\includegraphics[width=15cm]{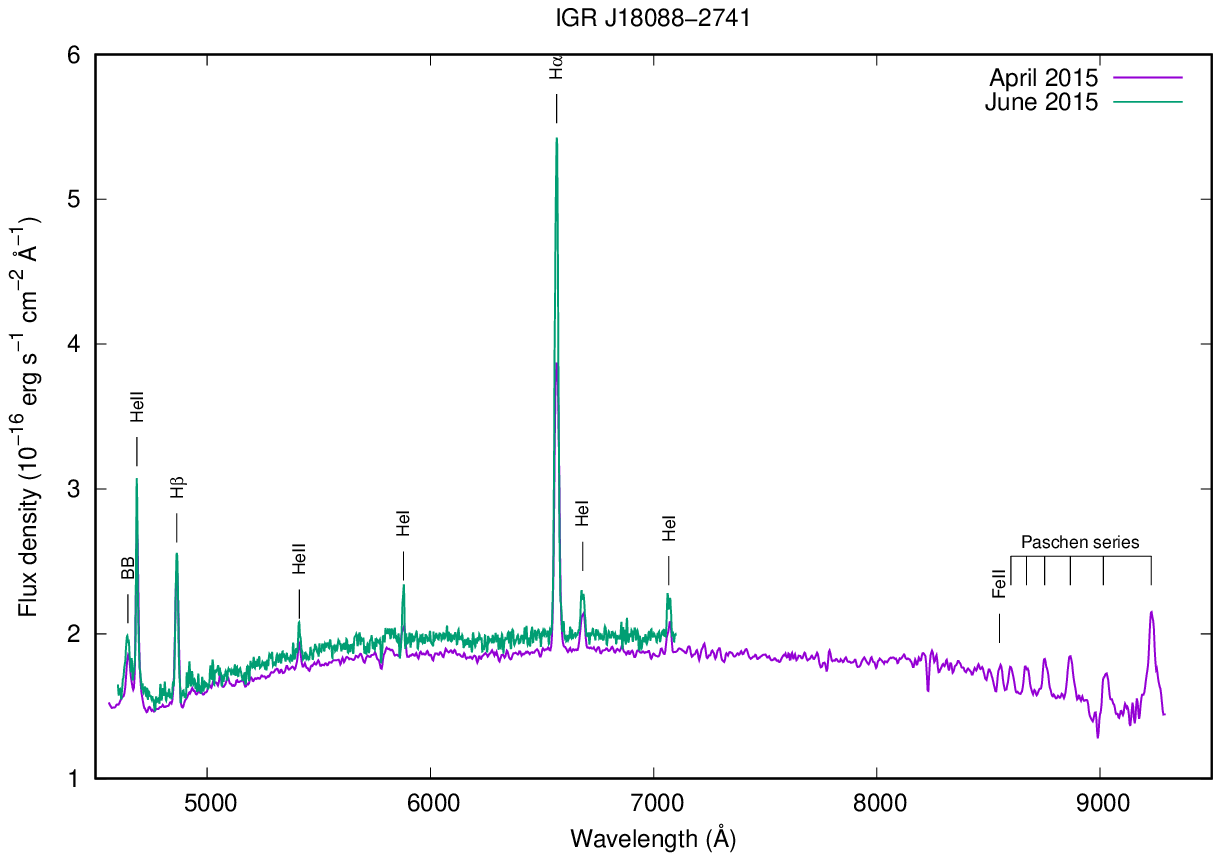}
\caption{\small Flux-calibrated FORS2 spectrum of \088\ observed in April 2015 (magenta) and June 2015 (green). All the detected emission lines
  are marked. The most prominent telluric absorption troughs were manually removed.}
\label{spec18088}
\end{center}
\end{figure*}

Comprehensive multi-wavelength observations are crucial to overcome this issue. This is the reason why several groups, including ours, are involved in large observational campaigns using multiple ground and space-based facilities with the goal of improving our view on hard X-ray emitters. So far, the majority of the new IGR sources that were identified are, as expected, active galactic nuclei (AGN). Nonetheless, an important result of the \intl\ surveys and subsequent multi-wavelength follow-ups have been the characterisation of a new population of peculiar persistent and transient supergiant X-ray binaries (SGXBs) in which a neutron star (NS) fed on the stellar winds of a blue supergiant through Bondi-Hoyle processes  \citep[see e.g.][]{2006Negueruelaa, 2008Tomsick, 2008Chaty, 2008Rahouia, 2008Rahouib, 2012Tomsick, 2012Bodagheeb, 2012Chaty, 2013Coleirob}. These discoveries significantly increased the number of known Galactic SGXBs and improved our understanding of their properties, although several unknowns remain, such as the origin of the erratic behaviour of transient SGXBs or the existence of a possible evolutionary link between the two populations. However, one of the most pressing issue is the absence of SGXBs hosting a black hole (BH) rather than a NS in the sample of newly-discovered sources. Such a phenomenon is puzzling, and it is not clear if it is due to formation and evolution processes \citep{2009Belczynski} or to the relatively low accretion power of BH-SGXBs, which makes them almost undifferentiated to isolated supergiant stars \citep[see, e.g.,][]{2004Zhang, 2014Casares, 2014Munar}. In the era of the first ever detection of gravitational waves from a system of two colliding stellar-mass BHs \citep{2016Abbott}, a likely progenitor of which was a BH-SGXB, finding out if we can expect such systems to exist in high numbers in our own Galaxy is crucial. 

Continuing with our ongoing campaign aiming at identifying newly-discovered hard X-ray sources, we report here on an optical spectroscopic study of three unidentified IGR sources, \088, \381, and \164, which follows the discovery of their soft X-ray counterparts \citep{2016Tomsick, 2016Tomsickb}. They were chosen because (1) they are located in the vicinity of the Galactic Plane; (2) their \chandra\ position was accurate enough to pinpoint their optical counterparts (see \autoref{imsources}); and (3) their X-ray spectra are hard and consistent with that expected from SGXB candidates. We added \9173, previously studied via soft X-ray and optical spectroscopy by \citet{2011Pavan} and \citet{2012Masetti}, respectively, because it was tentatively classified as an HMXB but doubts remain. We present the data and their reduction in Section~2, whereas Section~3 is devoted to their analysis. We discuss the outcomes in Section~4 and conclude in Section~5.
\begin{table*}
\caption{Optical lines in IGR~J18088$-$2741 2015 April and June FORS2 spectra. Note that June spectrum beyond 7100~\AA\ was contaminated by clouds and we decided not the measure the lines, which were however present. \label{speclines}}
\begin{center}
\begin{tabular}{l|c|ccc||ccc}
\hline\hline
&&\multicolumn{3}{c||}{April 2015}&\multicolumn{3}{c}{June 2015}\\
\hline
Element&${\lambda_{\rm \small c}}^a$&${\mathring{W}}^b$&FWHM$^{c}$&${F_{\rm line}}^d$&${\mathring{W}}$&FWHM&${F_{\rm line}}$\\
\hline
BB	   	&	4645	&$-7.2\pm0.4$	     &2010$\pm$135   &1.14$\pm$0.05 	&$-10.6\pm$0.6&$2732\pm199$&1.60$\pm$0.10\\
\ion{He}{ii}	&	4686	&$-12.6\pm$0.2	     &666$\pm$72   &1.86$\pm$0.06	&$-10.5\pm0.3$&$586\pm38$&$1.66\pm0.09$\\
H{$\beta$}	&	4861	&$-10.7\pm$0.3       &705$\pm$51   &1.65$\pm$0.04	&$-9.7\pm0.3$&$786\pm43$&$1.51\pm0.09$\\
\ion{He}{ii}    &	5409	&$-1.5\pm$0.2        &551$\pm$72   &0.26$\pm$0.03 	&$-1.6\pm0.2$&$524\pm75$&0.32$\pm$0.04\\
\ion{He}{i} 	&	5875	&$-3.0\pm$0.4	     &353$\pm$57   &0.55$\pm$0.08	&$-2.7\pm0.3$&$430\pm37$&$0.53\pm0.05$\\
H{$\alpha$}	&	6561	&$-26.7\pm$0.2       &652$\pm$45   &5.05$\pm$0.05	&$-32.4\pm0.1$&$769\pm37$&$6.51\pm0.05$\\
\ion{He}{i}	&	6678	&$-3.3\pm$0.2	     &804$\pm$92   &0.62$\pm$0.04	&$-3.1\pm0.3$&$825\pm74$&$0.62\pm0.07$\\
\ion{H}{i}	&	7065	&$-2.4\pm$0.2	     &699$\pm$74    &0.51$\pm$0.04 	&$-2.9\pm0.4$&$662\pm66$&$0.57\pm0.08$\\
\ion{H}{i}	&	8543	&$-2.8\pm$0.3	     &401$\pm$51   &0.45$\pm$0.04 	&--&--&--\\
\ion{H}{i}	&	8597	&$-2.9\pm$0.3	     &377$\pm$40   &0.46$\pm$0.05 	&--&--&--\\
\ion{H}{i}	&	8663	&$-3.2\pm$0.3	     &402$\pm$39   &0.51$\pm$0.05 	&--&--&--\\
\ion{H}{i}	&	8749	&$-4.4\pm$0.2	     &437$\pm$30   &0.70$\pm$0.04 	&--&--&--\\
\ion{H}{i}	&	8862	&$-5.4\pm$0.3	     &507$\pm$39   &0.84$\pm$0.04 	&--&--&--\\
\ion{H}{i}	&	9015	&$-6.3\pm$0.4	     &420$\pm$51   &0.72$\pm$0.05 	&--&--&--\\
\ion{H}{i}	&	9228	&$-13.4\pm$1.0       &520$\pm$107  &2.08$\pm$0.20 	&--&--&--\\
\hline
\end{tabular}
\end{center}
\raggedright{\hspace{2.5cm}{{$^a$}{Measured wavelength in \AA}\\}
\hspace{2.5cm}{{$^b$}{Equivalent widths in \AA}\\}
\hspace{2.5cm}{{$^c$}{Full-width at half-maximum in \kms, quadratically corrected for instrumental broadening}\\}
\hspace{2.5cm}{{$^d$}{Intrinsic line flux in units of $10^{-15}\,\,{\rm erg\,\,cm}^{-2}\,\,{\rm s}^{-1}$}\\}}
\end{table*}

\begin{figure}
\begin{center}
\begin{tabular}{c}
\includegraphics[width=7.8cm]{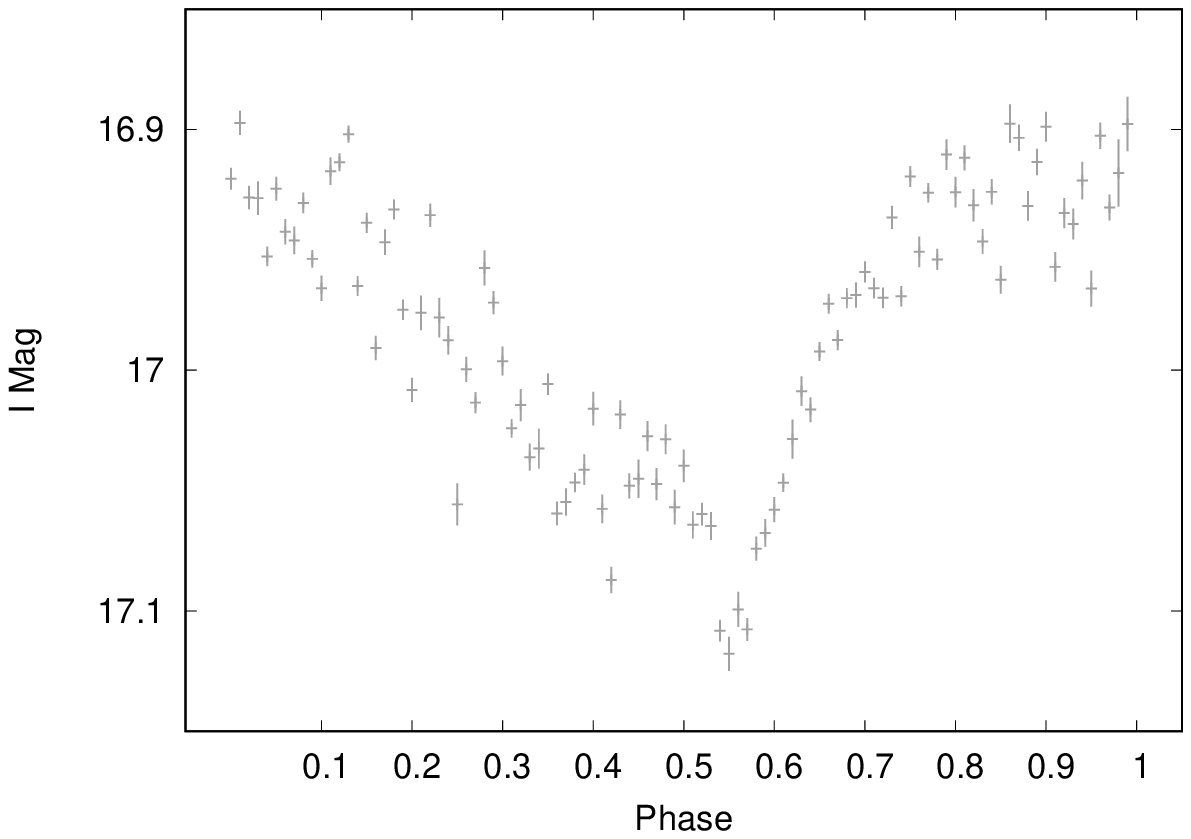}\\
\includegraphics[width=7.8cm]{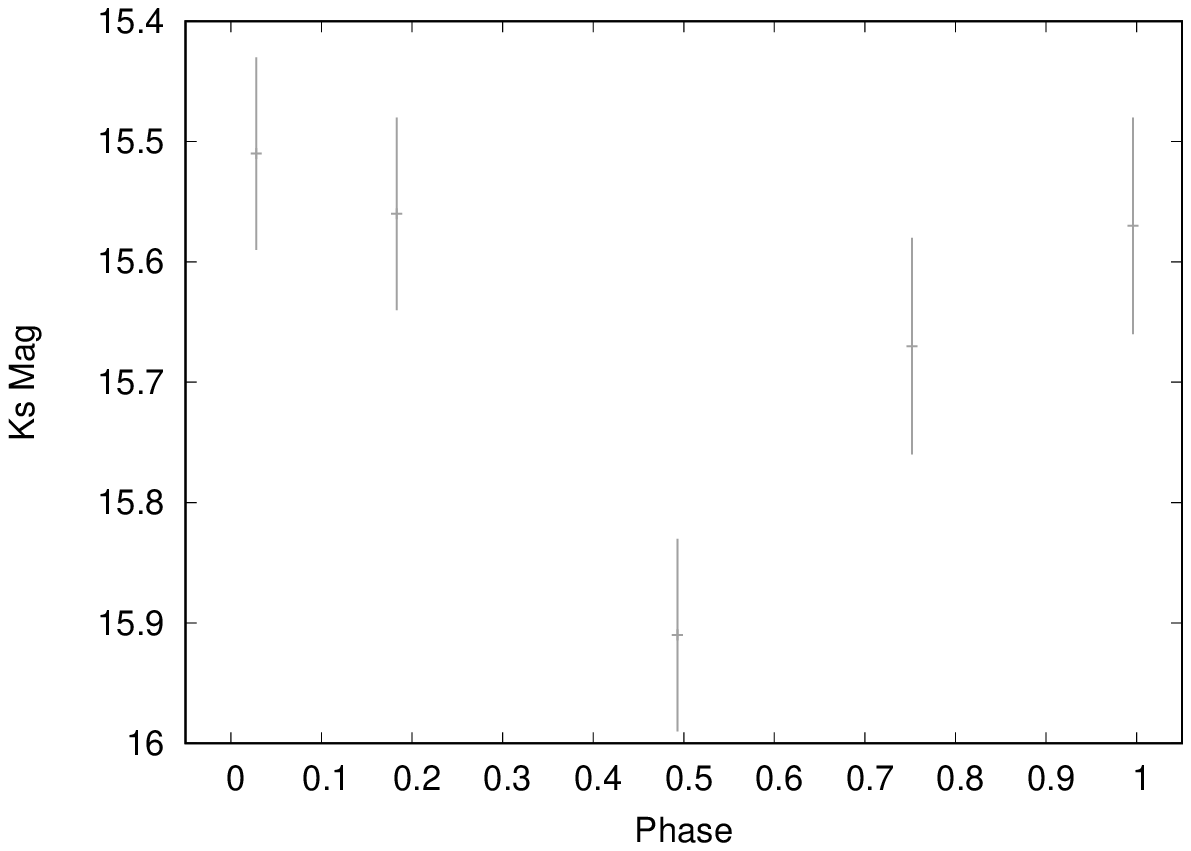}\\
\end{tabular}
\end{center}
\caption{\small {\bf Top}: Phase-folded $I$ band OGLE light curve of \088. {\bf Bottom} Phase-folded $K_{\rm S}$ band VVV light curve of \088. Note that the phase-folding is such that the time of maximum brightness occurs at phase 0 but this is arbitrary.}
\label{phase18088}
\end{figure}

\section{Observations and data reduction}

The dataset consists of (1) optical spectroscopy of the four sources with the FOcal Reducer/Low disperser Spectrograph 2 \citep[FORS2,][]{1998Appenzeller}, mounted on the UT1 Cassegrain focus at the ESO Very Large Telescope (VLT) at Cerro Paranal; and (2) soft X-ray spectroscopy of \381\ with the PN-CCD and MOS-CCD cameras of the European Photon Imaging Camera \citep[EPIC,][]{2001Struder, 2001Turner} on board \xmm.

\subsection{Optical spectroscopy}

On 2015 April 26, low-resolution spectroscopy of \088, \381, and \164\ 
was carried out with FORS2. The 300V and 300I grisms were used in combination with the GC435 and OG590 filters, respectively, for a total $4500-9300$~\AA\ spectral coverage. In both cases, the slit-width was set to 1\arcsec, giving a $R\sim600-700$ average spectral resolution.  Atmospheric conditions were relatively good, with a thin sky transparency, a seeing at 500~nm in the range 0\farcs6-0\farcs7, and an airmass always close to 1. Depending on the filters and photometric brightness of the sources, the exposure time of each individual frame ranged between 10 and 600~s and 2 exposures were taken per filter per source. 

We also performed medium-resolution spectroscopy of \088, \381, and \9173\ with FORS2 on 2015 June 27. We this time used the 600V and 600I grisms combined with the GC435 and OG590 filters, respectively, for a total $4500-9300$~\AA\ spectral coverage. The slit-width was set to 1\arcsec, giving a better $R\sim1400$ average spectral resolution. Atmospheric conditions were similar, with a thin sky transparency, a seeing at 500~nm in the range 0\farcs5-0\farcs6, and airmass between 1.3 and 1.6. Individual exposure times were set from 200 to 600~s and 2 exposures were taken per filter per source.

In both April and June 2015, the A0V spectro-photometric standard star CD-32~9927 was observed in similar conditions for flux-calibration. We reduced the data using the dedicated pipeline (v.~5.3.5) implemented in the ESO data reduction environment {\tt Reflex}~v.~2.6 \citep{2013Freudling}. It follows the standard steps for optical spectroscopy reduction and produces cleaned, background-subtracted, and wavelength-calibrated 2D spectroscopic images. We then used the routines {\tt apall}, {\tt
  standard}, {\tt sensfunc}, and {\tt calibrate} implemented in {\tt IRAF}~v.~2.16\footnote{ {\tt  IRAF} is distributed by the National Optical Astronomy Observatories, which are operated by the Association of Universities for Research in Astronomy, Inc., under cooperative agreement with the National Science Foundation.} to optimally extract the source and spectro-phometric standard star 1D spectra, compute the sensitivity function, and apply it to the sources spectra for flux calibration.

\subsection{\xmm\ observation of \381}

\begin{figure*}
\begin{center}
\begin{tabular}{cc}
\includegraphics[width=8cm]{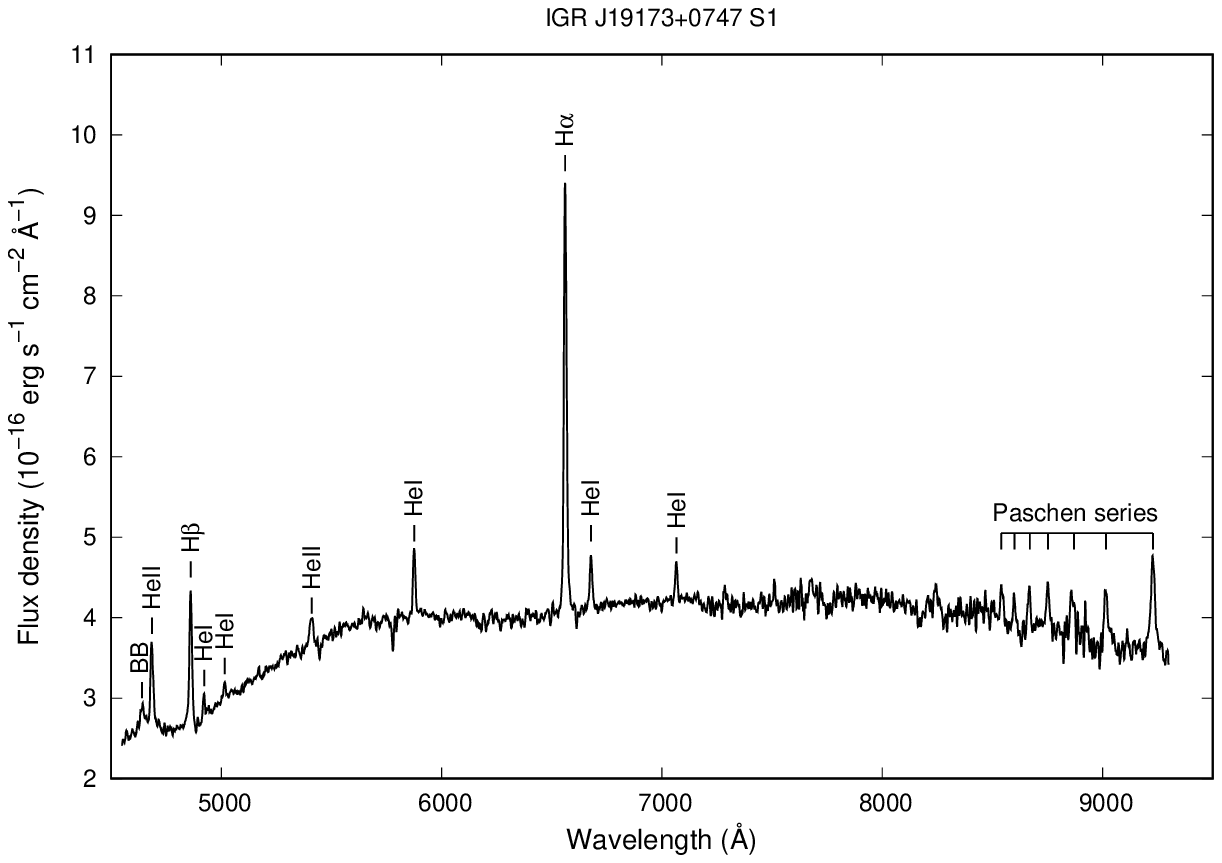}&\includegraphics[width=8cm]{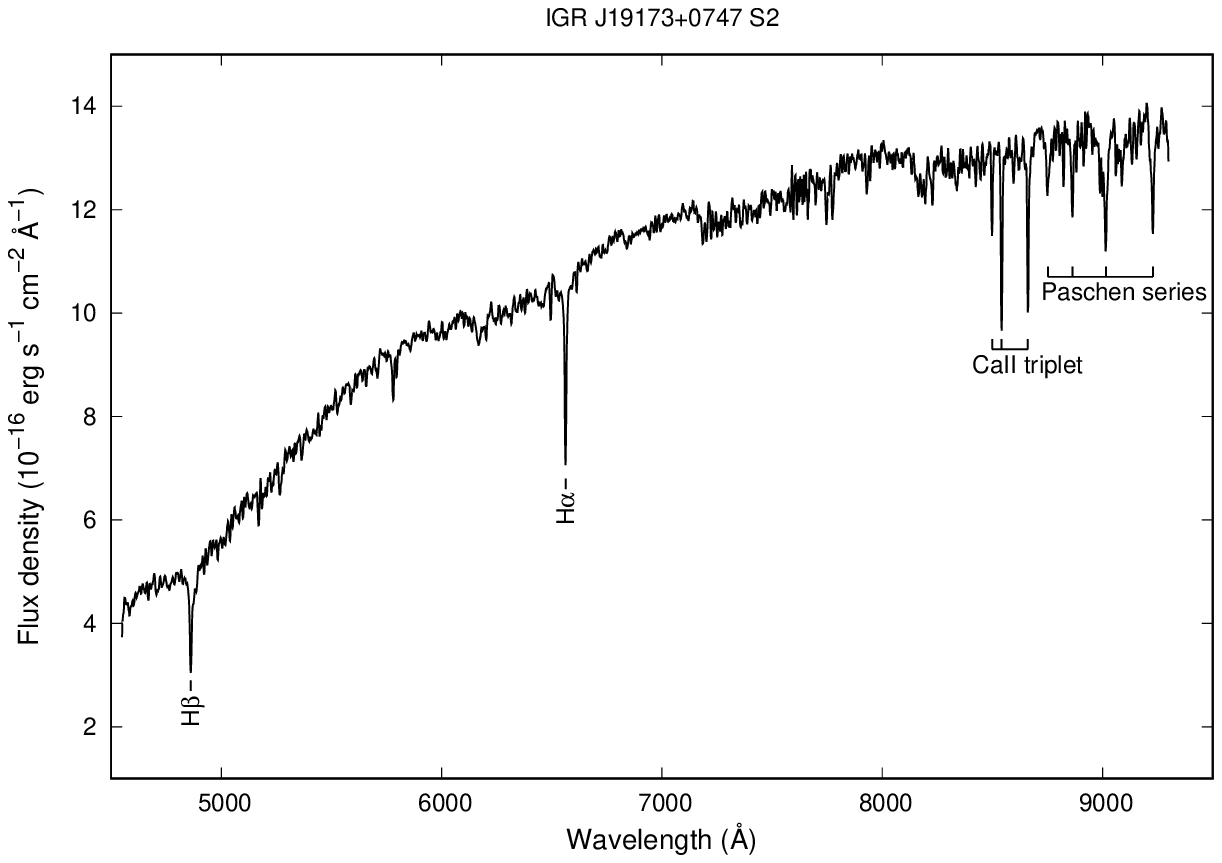}\\
\end{tabular}
\caption{Flux-calibrated FORS2 spectra of the two possible optical counterparts to \9173, a likely CV (S1, left) and an F3~V star (S2, right). All the detected lines are marked. The most prominent telluric absorption troughs were manually removed.}
\label{spec19173}
\end{center}
\end{figure*}

An observation of \381\ with \xmm\ occurred
on 2016 March 19 between 12.5\,h and 20.5\,h UT. The PN, MOS1, and
MOS2 instruments were all operated in Full Frame mode with the
medium blocking filter, yielding exposure times of 24, 28, and 28~ks,
respectively.  We previously identified \381\ with the
\chandra\ source CXOU~J183818.5--092552 \citep{2016Tomsick},
and we detect a source with a consistent position in the \xmm\ data.

We reduced the data using the \xmm\ Science Analysis
Software (SAS) v14.0.0 and the 2015 September 2 version of the
instrument calibration files.  We filtered the PN event list to
select good events using the \#XMMEA\_EP criterion and included
single and double events (``PATTERN<=4'').  For the MOS event
lists, we also selected good events using \#XMMEA\_EM but included
single through quadruple events (``PATTERN<=12'') as recommended
based on the SAS Data Analysis Threads\footnote{See http://www.cosmos.esa.int/web/xmm-newton/sas-threads}.

To extract the energy spectra, we used a circular region centred
on the source with a radius of 35\arcsec.  We made background
spectra by extracting events from rectangular regions of the detector
that are free of sources.  We binned the 0.3--12\,keV PN spectrum by
requiring a signal-to-noise ratio of at least 5 in each bin and
required a signal-to-noise $>$3 for the 0.3--10\,keV MOS spectra.
After background subtraction, we obtained PN, MOS1, and MOS2 count rates
of $0.248\pm 0.004$, $0.077\pm 0.002$, and $0.083\pm 0.002$~count/s,
respectively.

\section{Results and analysis}

In the following, we summarise the main properties of the four sources as they were known before the present study and we give some insights on the most likely nature of their optical counterparts.

\subsection{IGR~J18088$-$2741}

\088\ was first reported in \citet{2012Krivonos}, and \citet{2016Tomsick} later identified it as  CXOU~J180839.8$-$274131 with \chandra. The fit to its combined \chandra\ and \intl\ spectra with a cut-off power law yields a very hard $\Gamma\approx -1.5$ and  $E_{\rm fold}\approx 4.8$~keV, whereas the column density is $\nhe < 7 \times 10^{21}$~\cm2. The authors also find a 800 to 950 s modulation in its \chandra\ light curve (although this was based on coverage of only five cycles of the modulation) and the 0\farcs74 \chandra\ positional accuracy led to an unambiguous association of \088\ with VVV~J180839.77$-$274131.7 \citep[Vista Variables in the Via Lactea catalogue,][]{2010Minniti}. The reported magnitudes of the source are $Z=16.77\pm0.06$, $Y=16.65\pm0.07$, $J=16.09\pm0.06$, $H=15.82\pm0.08$, and $K_{\rm S}=15.67\pm0.09$, which yields a flux density of about $1.7\times10^{-16}$~\ergcms\ at 0.88~\mic\ compared to about $1.5\times10^{-16}$~\ergcms\ for the continuum level of our first FORS2 spectrum. This infrared counterpart is also that of OGLE-BLG-RRLYR-14363, a variable star with a 0.28 days orbital period listed in the Optical Gravitational Lensing Experiment Catalogue of Variable Stars \citep[OGLE-III,][]{2011Soszynski}. We note that the Vista source is also variable and a comparison of the phase-folded OGLE/$I$ and VVV/$K_{\rm S}$ light curves (\autoref{phase18088}) unambiguously shows that these variations are due to orbital modulation.bb

Based on its very hard cut-off power law X-ray spectrum, the possible presence of pulse period, as well as the very short orbital period, \citet{2016Tomsick} classified \088\ as a likely intermediate polar (IP), i.e. a magnetised cataclysmic variable (CV); our FORS2 optical spectra concur with this statement (see \autoref{spec18088}; note that we do not display the red part of the June spectrum as it was contaminated by passing clouds and many narrow absorption troughs not related to the source were present). Indeed, the optical light is dominated by strong emission signatures of \ion{H}{i} (Balmer and Paschen series) and \ion{He}{i}, which, along with the detection of the Bowen Blend around 4640~\AA\ and \ion{He}{ii} at 4686~\AA, point towards the presence of an irradiated accretion disc. Moreover, if the continuum and most of the emission lines are constant between April and June observations, the Bowen Blend and \ha\ are a notable exception, with a flux increase between the two epochs. This may hint at orbital modulations and a brightening of the hot spot of the accretion disc.  

Centred at 5779~\AA, we also report a Diffuse Interstellar Band (DIB) whose equivalent width is known to correlate well with the extinction along the line-of-sight of the sources for which it is detected \citep{1994Jenniskens}. Averaging out DIB5779 equivalent width values found in April and June spectra, i.e. $\mathring{W}=0.65\pm0.06$~\AA\ and $\mathring{W}=0.53\pm0.11$~\AA, respectively, we find that \088\ suffers from an ISM extinction $E(B-V)=0.96\pm0.10$ along its line-of-sight, i.e. $\Ave=2.98\pm0.31$ for $R_{\rm V}=3.1$. Using the 3D extinction map built in \citet{2006Marshall}, we find that it is consistent with a distance to the source $D \ge 9$~kpc. The more recent 3D extinction map of \citet{2015Green} gives a similar value, with $D \ge 8$~kpc. Moreover, correcting both spectra using the extinction law given in \citet{1999Fitzpatrick}, we estimate that the Balmer decrement \ha/\hb\ is about 1.02 and 1.47 in April and June, respectively, i.e. consistent with unity as expected for Balmer lines originating from the optically thick accretion disc in CVs \citep{1980Williams, 2016Tomsick}.
\begin{table}
\caption{Optical lines in \9173-S1 June 2015 FORS2 spectra. \label{specline19173}}
\begin{center}
\begin{tabular}{l|c|ccc}
\hline\hline
Element&${\lambda_{\rm \small c}}^a$&${\mathring{W}}^b$&FWHM$^{c}$&${F_{\rm line}}^d$\\
\hline
BB	   &	4641	&$-7.3\pm0.8$&$1974\pm307$&1.81$\pm$0.20\\
\ion{He}{ii}&	4685	&$-10.8\pm0.6$&$841\pm74$&$2.70\pm0.13$\\
H{$\beta$}	&	4861	&$-9.0\pm0.3$&$731\pm41$&$2.39\pm0.08$\\
\ion{He}{i}&	4921	&$-1.0\pm0.2$&$423\pm83$&0.27$\pm$0.05\\
\ion{He}{i}&	5015	&$-1.1\pm0.2$&$691\pm139$&0.32$\pm$0.05\\
\ion{He}{ii}&	5409	&$-1.7\pm0.2$&$935\pm133$&0.63$\pm$0.08\\
\ion{He}{i} &	5876	&$-3.2\pm0.3$&$606\pm38$&$1.29\pm0.09$\\
H{$\alpha$}	&	6561	&$-19.1\pm0.1$&$562\pm26$&$7.92\pm0.06$\\
\ion{He}{i}	&	6678	&$-2.0\pm0.2$&$462\pm40$&$0.81\pm0.07$\\
\ion{H}{i}	&	7065	&$-1.5\pm0.2$&$440\pm48$&$0.64\pm0.08$\\
\ion{H}{i}	&	8543	&$-1.7\pm$0.4&493$\pm$62   &0.69$\pm$0.08\\
\ion{H}{i}	&	8599	&$-1.1\pm$0.2&286$\pm$40   &0.41$\pm$0.07\\
\ion{H}{i}	&	8667	&$-2.0\pm$0.3&424$\pm$43   &0.75$\pm$0.07\\
\ion{H}{i}	&	8752	&$-2.2\pm$0.4&467$\pm$47   &0.84$\pm$0.08\\
\ion{H}{i}	&	8862	&$-3.7\pm$0.4&809$\pm$90   &1.41$\pm$0.15\\
\ion{H}{i}	&	9016	&$-4.1\pm$0.4&635$\pm$60   &1.52$\pm$0.12\\
\ion{H}{i}	&	9228	&$-6.0\pm$0.9&616$\pm$41  &2.21$\pm$0.20\\
\hline
\end{tabular}
\end{center}
\raggedright{\hspace{0.5cm}{{$^a$}{Measured wavelength in \AA}\\}
\hspace{0.5cm}{{$^b$}{Equivalent widths in \AA}\\}
\hspace{0.5cm}{{$^c$}{Full-width at half-maximum in \kms, quadratically corrected for \\
\hspace{0.6cm}{instrumental broadening}\\}}
\hspace{0.5cm}{{$^d$}{Intrinsic line flux in units of $10^{-15}\,\,{\rm erg\,\,cm}^{-2}\,\,{\rm s}^{-1}$}\\}}
\end{table}

\subsection{IGR~J19173$+$0747}

\9173\ was first reported in \citet{2011Pavan}, in which the authors investigated the presence of unknown sources in the vicinity of AX~J1910.7+0917 \intl\ field-of-view. They performed a fit to its 17--80~keV ISGRI spectrum and found that it was best-described by a $\Gamma=3.3_{-0.7}^{+0.9}$ power law. The follow-up observations with the X-Ray Telescope \citep[XRT,][]{2005Burrows} on board the \sw\ satellite led to the identification of its soft X-ray counterpart, the 0.5--10~keV spectrum of which was best-fit with a $\Gamma=0.6\pm0.2$ power law, pointing towards a break in the 10--20~keV energy range and/or variability. Moreover, the 3\farcs4 positional accuracy of the \sw/ observation allowed \citet{2011Pavan} to associate \9173\ with USNOB1.0~0977$-$0532587 and 2MASS~J19172078+0747506. \citet{2012Masetti} subsequently reported on optical spectroscopy of the optical counterpart and concluded that \9173\ was a likely HMXB based on its X-ray properties and the sole detection of an \ha\ emission line in its optical spectrum. Our results however contradict this statement.

Indeed, as shown in the top-right panel of \autoref{imsources}, which displays the $V$ band FORS2 acquisition image of \9173, what is classified as a single point source in both USNO and 2MASS catalogues actually consists of two objects. They are separated by about 2.2\arcsec\ and are well within the \sw\ error circle, meaning that both are potential optical counterparts to \9173. We labelled the northern source S1 and the southern one S2 and \autoref{spec19173} displays their FORS2 spectra; we also report the measurements of S1's detected lines in \autoref{specline19173}. S1 spectrum strongly resembles that of \088, with a wealth of \ion{H}{i} and \ion{He}{i} emission lines as well as the presence of the Bowen Blend and \ion{He}{ii}~$\lambda4686$. This similarity, combined with the fact that its 0.5-80~keV spectrum is likely best-described by a hard cut-off power law, suggests that S1 is the optical counterpart to a CV and that we observe the optically thick accretion disc. This statement is further strengthens by the fact that the Balmer decrement is \ha/\hb$\approx$0.96, i.e. close to unity as expected for CVs. We assessed this decrement using the measured ISM extinction along S1 line-of-sight, $\Ave=3.35\pm0.50$, derived from the equivalent width of DIB5779, $\mathring{W}=0.70\pm0.09$~\AA, using the relationship given in \citet{1994Jenniskens}. Incidentally, this extinction yields a distance between 3 and 5~kpc using the 3D extinction map given in \citet{2006Marshall}, and between 1.5 and 2.9~kpc using the work of \citet{2015Green}. 

\begin{figure*}
\begin{center}
\includegraphics[width=15cm]{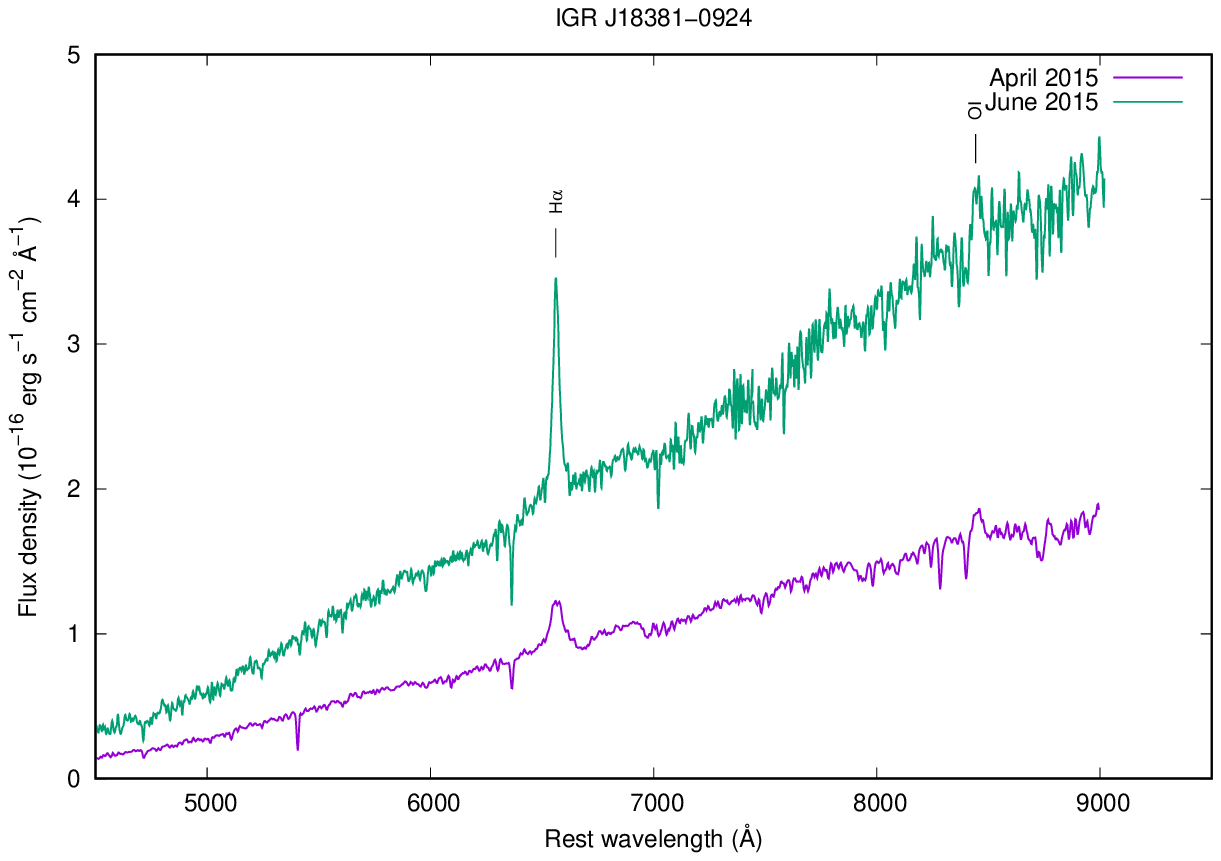}
\caption{\small Rest-frame flux-calibrated FORS2 spectrum of \381\ observed in April 2015 (magenta) and June 2015 (green). All the detected emission lines
  are marked. The most prominent telluric absorption troughs were manually removed.}
\label{spec18381}
\end{center}
\end{figure*}

In contrast to S1, S2 spectrum is dominated by absorption signatures of \ha, \hb, Paschen lines at 8752~\AA, 8864~\AA, 9017~\AA\ and 9230~\AA, as well as the \ion{Ca}{ii} triplet at 8498~\AA\, 8542~\AA, and 8662~\AA. Comparing it to the optical spectra provided in the STELIB library \citep{2003Leborgne}, we identify S2 as a likely F3~V star. DIB5779 is also present in its spectrum and we measure its equivalent width to be $\mathring{W}=0.75\pm0.08$~\AA, which results in an ISM extinction $\Ave=3.60\pm0.47$ and a distance between 3 and 5~kpc. This means that S2 suffers from the same ISM extinction and is located at the same distance as S1, which may hint at both belonging to the same group of stars. 

\subsection{IGR~J18381$-$0924}

\begin{table*}
\caption{Red-shifted optical emission lines in \381\ April and June 2015 FORS2 spectra. \label{speclines_18381}}
\begin{center}
\begin{tabular}{l|c|ccc||ccc}
\hline\hline
&&\multicolumn{3}{c||}{April 2015}&\multicolumn{3}{c}{June 2015}\\
\hline
Element&${\lambda_{\rm \small c}}^a$&${\mathring{W}}^b$&FWHM$^{c}$&${F_{\rm line}}^d$&${\mathring{W}}$&FWHM&${F_{\rm line}}$\\
\hline
H{$\alpha$}	&	6767	&$-26.4\pm1.3$       &2614$\pm$224   &2.41$\pm$0.13	&$-27.3\pm0.6$&$1926\pm102$&$4.87\pm0.12$\\
\ion{O}{i}	&	8708	&$-4.8\pm$1.2	     &1573$\pm$428   &0.81$\pm$0.19 &$-4.8\pm0.7$&$1571\pm298 $&$1.77\pm0.25$\\
\hline
\end{tabular}
\end{center}
\raggedright{\hspace{2.6cm}{{$^a$}{Measured wavelength in \AA}\\}
\hspace{2.6cm}{{$^b$}{Equivalent widths in \AA}\\}
\hspace{2.6cm}{{$^c$}{Full-width at half-maximum in \kms, quadratically corrected for instrumental broadening}\\}
\hspace{2.6cm}{{$^d$}{Intrinsic line flux in units of $10^{-15}\,\,{\rm erg\,\,cm}^{-2}\,\,{\rm s}^{-1}$}\\}}
\end{table*}

\begin{figure}
\begin{center}
\includegraphics[width=8cm]{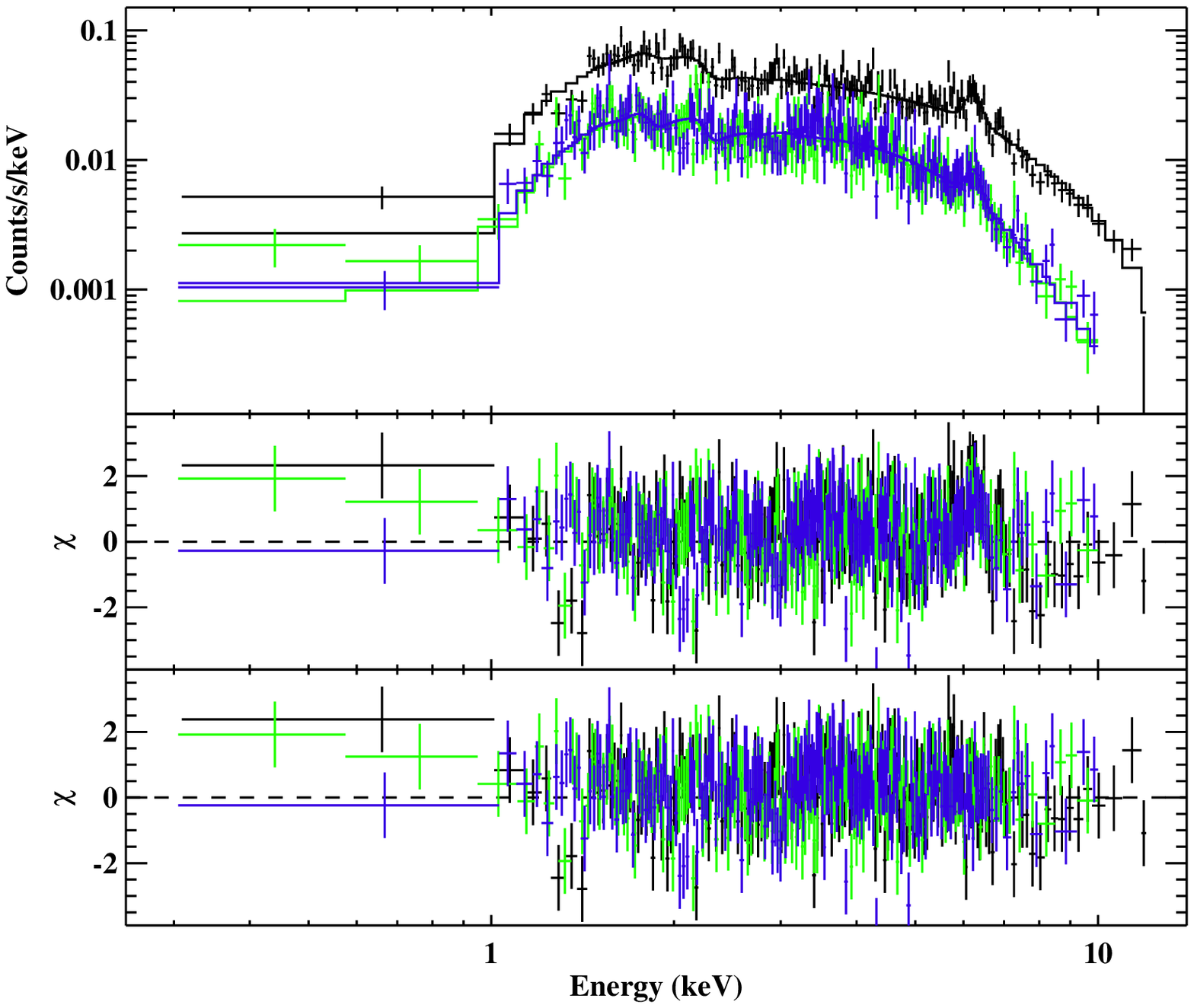}
\caption{\small {\bf (Top)} PN (black), MOS1 (green) and MOS2 (blue) \xmm\ best-fit spectra of \381. {\bf (Middle)} Residuals for an absorbed power law, pointing towards the presence of an emission line between 6 and 7~keV. {\bf (Bottom)} Residuals for an absorbed power law combined with a Gaussian centred around 6.24~keV.}
\label{xfitfigure18381}
\end{center}
\end{figure}

Similarly to \088, \381\ was first reported in \citet{2012Krivonos} and studied in more details in \citet{2016Tomsick} via both \chandra\ and \intl\ X-ray spectroscopy. The authors showed that the source had a relatively hard spectrum consistent between ACIS-I and ISGRI, with a best-fit power law $\Gamma=1.4\pm0.1$ and $\nhe=3.8_{-0.7}^{+0.9}\times 10^{22}$~\cm2. The 0\farcs74 \chandra\ pointing accuracy allowed for the identification of two possible near-infrared counterparts to the source in the UKIDSS Galactic Plane Survey \citep[UGPS,][]{2008Lucas}, UGPS~J183818.58-092552.9 and UGPS~J183818.59-092551.8, both classified as galaxies with reported $K$ magnitudes $13.035\pm0.003$ and $12.833\pm0.002$. 

In the optical however, as shown in \autoref{imsources}, only UGPS J183818.59-092551.8 is detected. Our FORS2 spectra, obtained in April and June 2015 (\autoref{spec18381}, magenta and green, respectively) confirm its classification as extragalactic. Indeed, we only detect red-shifted emission signatures of \ha\ and \ion{O}{i} at 8446~\AA, which yields a red shift $z=0.031\pm0.002$ (\autoref{speclines_18381}). The source is also variable and the continuum is twice as bright in June compared to April. Measurements show that the two emission lines, relatively broad with FWHMs of about 2000--2600~\kms\ for \ha\ and 1600~\kms\ for \ion{O}{i}, also brightened at constant equivalent widths, pointing towards a common origin with the continuum.  

\begin{figure}
\begin{center}
\includegraphics[width=8cm]{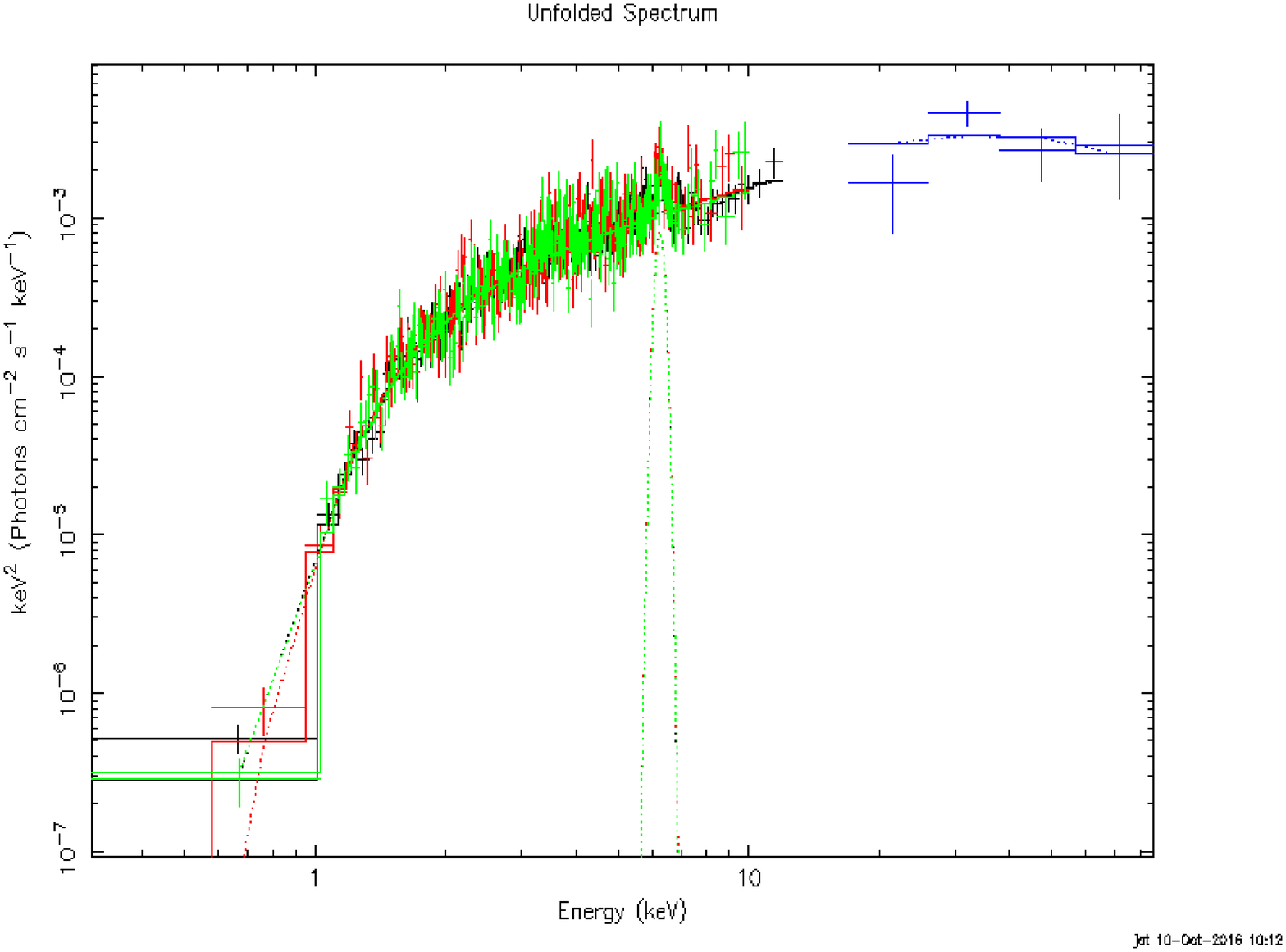}
\caption{\small Best fit of the \xmm/PN and \intl/ISGRI joint spectrum of \381. The model used in a cutoff power law combined to a red-shifted iron line, both modified by photoelectric absorption.}
\label{xfitfigure18381xmmint}
\end{center}
\end{figure}

\begin{table}
\begin{center}
\caption{Best-fit parameters to the
   \381\ \xmm/PN+MOS (left) as well as \xmm/PN+MOS and \intl/ISGRI (right) spectra with {\sc constant$\times$tbabs$\times$(pegpwrlw+zgauss}). \label{xfitparam18381}}
\begin{tabular}{lcc}
\hline\hline
\nh$^{a}$&$2.21\pm0.16$&$2.22\pm0.16$\\
$\Gamma^{b}$&$1.19\pm0.07$&$1.19\pm0.07$\\
$E_{\rm iron}^{c}$&$6.24^{+0.05}_{-0.09}$&$6.24^{+0.05}_{-0.09}$\\
$\sigma_{\rm iron}^{d}$&$0.15_{-0.07}^{+0.18}$&$0.15_{-0.07}^{+0.19}$\\
$z^{e}$&$0.026_{-0.008}^{+0.016}$&$0.026_{-0.008}^{+0.016}$\\
$C_{\rm MOS1}^{f}$&$1.00\pm0.05$&$1.00_{-0.05}^{+0.04}$\\
$C_{\rm MOS2}^{g}$&$0.96\pm0.04$&$0.96\pm0.04$\\
$C_{\rm ISGRI}^{h}$&--&$0.64_{-0.19}^{+0.21}$\\
$F_{1-10}^{i}$&$2.3\pm0.08$&--\\
$\chi_{\rm r}^2$ (d.o.f)&1.00 (582)&1.01 (584)\\
\hline
\end{tabular}
\end{center}
\raggedright{\hspace{2.6cm}{{$^a$}{Column density in units of $10^{22}$~cm$^{-2}$}}\\
\hspace{2.6cm}{{$^b$}{Power law photon index}}\\
\hspace{2.6cm}{{$^c$}{Iron line energy (keV)}}\\
\hspace{2.6cm}{{$^d$}{Iron line width (keV)}}\\
\hspace{2.6cm}{{$^e$}{Redshift}}\\
\hspace{2.6cm}{{$^f$}{MOS1 multiplicative constant}}\\
\hspace{2.6cm}{{$^g$}{MOS2 multiplicative constant}}\\
\hspace{2.6cm}{{$^h$}{ISGRI multiplicative constant}}\\
\hspace{2.6cm}{{$^i$}{1-10~keV absorbed flux in units of \\
\hspace{2.6cm} {$10^{-12}$~\ergcms}}}}\\
\end{table}

\begin{figure*}
\begin{center}
\begin{tabular}{cc}
\includegraphics[width=8cm]{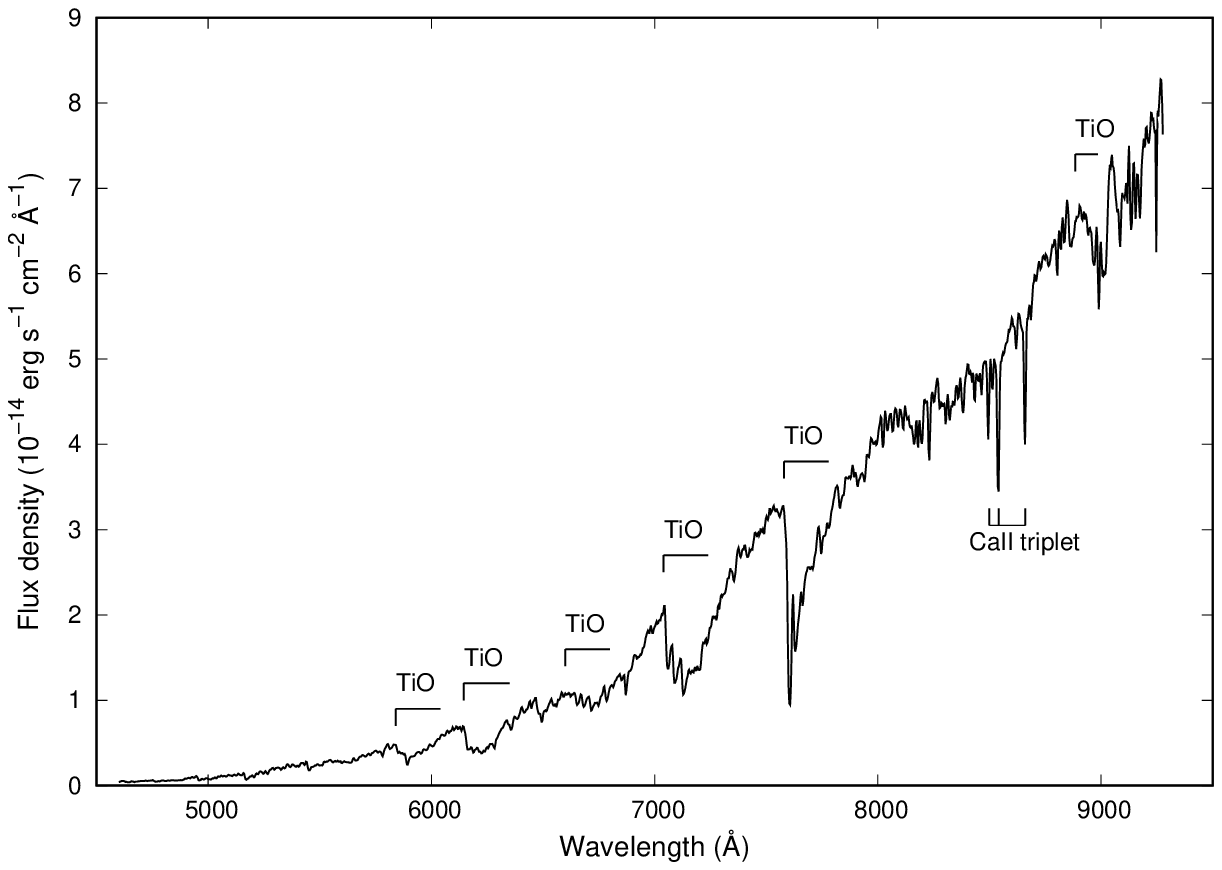}&\includegraphics[width=8cm]{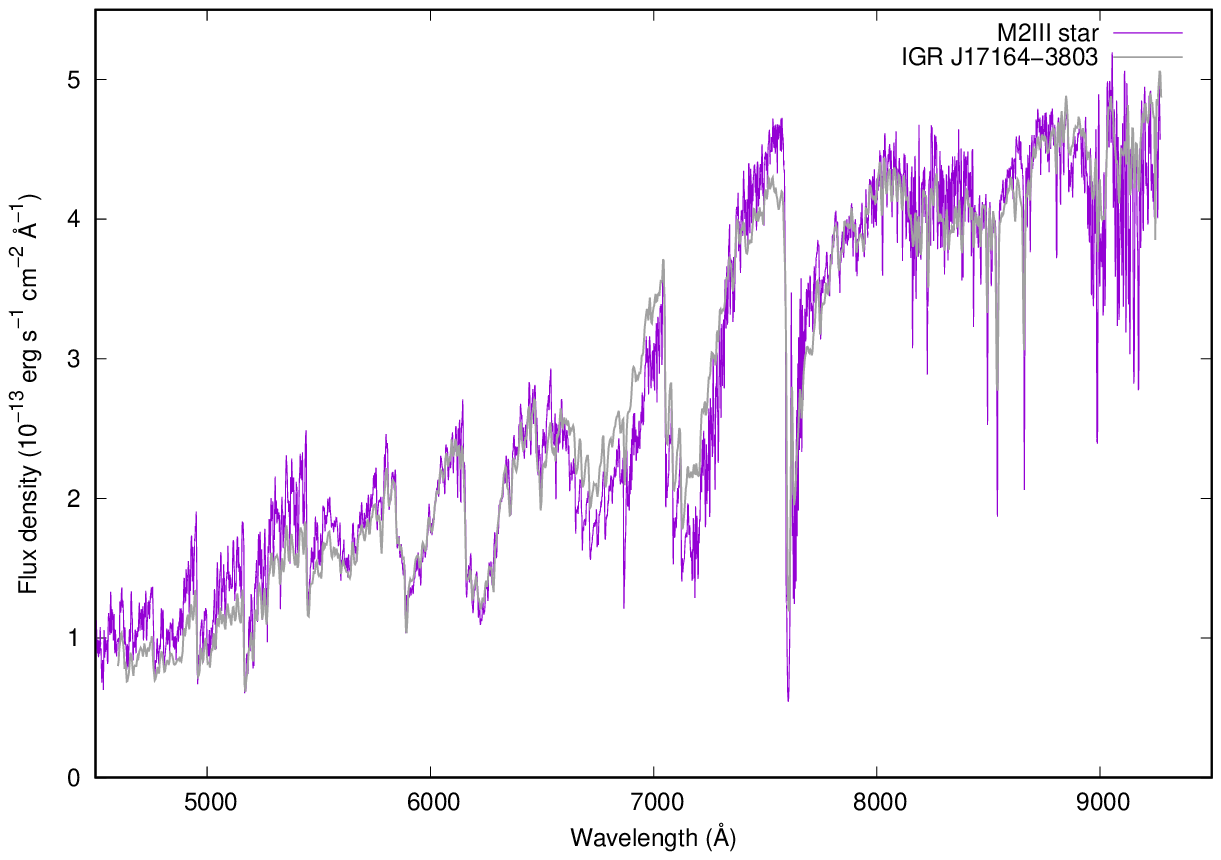}\\
\end{tabular}
\caption{{\bf Left:} flux-calibrated FORS2 spectrum of USNOB1.0~0519$-$0593639, the tentative optical counterpart to \164. {\bf Right:} Comparison of USNOB1.0~0519$-$0593639 FORS2 spectrum, corrected from an $\Ave\approx4.6$ ISM extinction, with that of a very weakly-extincted M2 giant star.}
\label{spec17164}
\end{center}
\end{figure*}

DIB5779 is also present, and its equivalent width $\mathring{W}=1.20\pm0.15$~\AA\ results in an extinction along the line-of-sight of the source $\Ave=5.75\pm0.85$. We note that this extinction, equivalent to $\nhe=(1.27\pm0.19)\times10^{22}$~\cm2\ using the relationship given in \citet{2009Guver}, is consistent with the total Galactic extinction of about $1.25\times10^{22}$~\cm2\ as obtained with the tool provided in the \chandra\ website\footnote{http://cxc.harvard.edu/toolkit/colden.jsp}. 
\newline

In the X-ray domain, similarly to what was done for the \chandra\ spectrum \citep{2016Tomsick}, we fit the PN, MOS1, and MOS2 \xmm\ spectra of \381\ with an absorbed power law. In modelling the absorption, we assumed \cite{2000Wilms} abundances and \cite{1996Verner} cross sections. We also included a multiplicative constant factor to account for possible normalisation differences between instruments. With a PN constant fixed to 1.0, we find $1.00\pm 0.05$ and $0.96\pm 0.04$ (90\% confidence errors) for MOS1 and MOS2, respectively, indicating that there are no significant normalisation differences. With the model {\sc Constant$\times$Tbabs$\times$Pegpwrlw}, we obtain $\chi^{2}/\nu = 635/585$ and see significant residuals between 6 and 7\,keV (see \autoref{xfitfigure18381}, middle panel, for the residuals). Adding a Gaussian emission line improves the fit to $\chi^{2}/\nu = 582/582$, and the central energy is $6.24^{+0.05}_{-0.09}$~keV, pointing towards a red-shifted iron line (see \autoref{xfitfigure18381}, top and bottom panels, for the best-fit and residuals, respectively, as well as \autoref{xfitparam18381}, second column, for the best-fit parameters). If we assume a rest energy of 6.4~keV, i.e. the presence of neutral iron as expected for an AGN, the inferred red shift is $z = 0.026^{+0.016}_{-0.008}$, which is consistent with the value measured from the optical spectrum of UGPS J183818.59-092551.8, proving that it is the proper optical counterpart to \381. The width of the line is $\sigma = 0.15^{+0.18}_{-0.07}$~keV, and the equivalent width is $280^{+130}_{-80}$~eV. The spectral parameters also indicate a photon index $\Gamma = 1.19\pm 0.07$ and a column density of $(2.21\pm 0.16)\times 10^{22}$~cm$^{-2}$. The latter is in excess of the total ISM extinction along the line-of-sight of the source as well as that derived from DIB5779, which hints at the presence of a local absorbing medium. We also note that the value derived with \xmm\ is lower than that obtained with \chandra, i.e. $(4\pm1)\times 10^{22}$~cm$^{-2}$, pointing towards variability. Correcting \381\ spectrum from this extinction, we measure an unabsorbed 1--10~keV flux of $(2.70\pm 0.08)\times 10^{-12}$~\ergcms, whereas the absorbed 1--10~keV flux is $(2.30\pm 0.06)\times 10^{-12}$~\ergcms. 

Finally, we combined our PN and MOS data with the ISGRI one and fit the joint spectrum with the same model. The best fit is displayed in \autoref{xfitfigure18381xmmint} and the best-fit parameters are listed in the third column of \autoref{xfitparam18381}. We obtain the very same results, proving that the \xmm\ and \intl\ data are consistent. However, the ISGRI multiplicative constant compared to PN, $0.64_{-0.19}^{+0.21}$, is quite low, which likely stems fron the variability of the X-ray emission.

\subsection{IGR~J17164$-$3803}

\164\ was detected by \intl\ with a 4\farcm2 positional accuracy \citep{2012Krivonos} and we later observed it with ACIS on \chandra. We find six \chandra\ sources within the \intl\ error circle \citep[see Table~5 in][]{2016Tomsick}, five of them being very faint with 0.5--10~keV count rates between $1.0\times10^{-3}$ and $2.4\times10^{-3}$ count/s. Although we cannot rule out any of them as the actual soft X-ray counterpart to \164, the latter would have to be either very hard, very absorbed, or variable to be that faint in the ACIS energy range \citep[see Table~4 in][]{2016Tomsick}. In contrast, the sixth source, which we consider as the most likely counterpart in the following, is brighter, with $5.7\times10^{-3}$ count/s, and is also the second closest to the \intl\ position, with RA=17$^h$~16$^m$~26.97$^s$ and DEC=$-38^\circ$ 00\arcmin\ 07.8\arcsec\ (Eq. J2000) for a 0\farcs73 uncertainty. We fit its 0.5--10~keV spectrum with anl absorbed power law, again for the \citet{2000Wilms} abundances and \citet{1996Verner} cross-sections, and we derive $\nhe=1.0_{-1.0}^{+2.4}\times 10^{22}$~\cm2\ and $\Gamma=1.0_{-1.0}^{+1.3}$, which points towards a relatively hard X-ray spectrum. 

The accurate soft X-ray position allowed us to pinpoint its bright low energy counterpart, which we unambiguously associate with USNOB1.0~0519$-$0593639. The left panel of \autoref{spec17164} displays its FORS2 spectrum, which consists in a red continuum dominated by TiO molecular troughs as well as the \ion{Ca}{ii} triplet in absorption. Overall, it resembles that of an early M giant star, and a comparison with several observed templates provided by \citet{2003Leborgne} shows that it is very likely an M2~III star with an approximate ISM extinction $\Ave\approx4.6$ (\autoref{spec17164}, right panel). We note that this value is roughly on par with the 3.5 to 5.5 estimate we find from DIB5779 equivalent width, the large range being due to the low resolution of our April spectrum. It also points towards USNOB1.0~0519$-$0593639 to be roughly located at 3 to 4~kpc.

\section{Discussion}

We find that (1) \088\ optical spectrum is that of an optically thick accretion disc; (2) the same can be said for one of the two possible counterparts of \9173, the other one being an F3~V star; (3) \381\ is a variable extragalactic source with $z\approx0.031$ red-shifted \ha\ and \ion{O}{i}~$\lambda8446$ emission lines; and (4) \164\ optical counterpart is an M2~III star. 

Among the four sources, \088\ is that with the most unambiguous identification. Besides its optical spectrum, its other properties, i.e. an X-ray emission best-described by a hard cut-off power law, the presence of pulsations, as well as that of orbital modulations leave little doubts that the source is a magnetic CV, as already suggested in \citet{2016Tomsick}. 

Likewise, the $z\approx0.031$ red shift and hard \xmm\ continuum suffering from a column density slightly in excess of the total Galactic extinction along its line-of-sight point towards \381\ being an obscured AGN. Based on the presence of broad \ha\ and \ion{O}{i}~$\lambda8446$ emission signatures, we tentatively classify the source as a type 1 galaxy in which we have a direct view of the broad emission line region (BELR).  We note that \ion{O}{i} is typical of a relatively dense region within the BELR with $n_{\rm h}\sim10^{12}$~cm$^{-3}$ in which Ly$\beta$ resonance fluorescence and collisional processes are important \citep[see, e.g.,][]{1980Grandi, 2002Ardila, 2007Matsuoka}. At this relatively low red shift, and taking into account the absence of a broad \hb\ component, this would suggest a Seyfert 1.9 nature \citep{1981Osterbrock}. It is however puzzling that no narrow forbidden emission lines are detected, as expected in Seyfert galaxies \citep[see, e.g.,][]{2013Lee}, but this may be due to the low X-ray luminosity of the source, i.e. $6.7\times10^{42}$~\ergs, which we estimate from our X-ray fit between 1 and 10~keV. Moreover, the strong optical continuum variability, i.e. a factor 2 increase in 2 months, points towards the presence of an irradiated accretion disc, similarly to what is seen for NGC~5548 \citep[see, e.g.,][]{2014Mchardy}.

Unlike the two previous sources, which have well-defined optical counterparts, \9173\ \sw\ position is consistent with two objects. S1 exhibits the optical spectrum of an optically thick accretion disc with a flat Balmer decrement, which is usually found in CVs. The best-fit to the X-ray spectrum of the source, i.e. a hard cut-off power law, is also consistent with a CV and this is the reason why we favour S1 as the optical counterpart of \9173. We however cannot completely rule out S2, even if it is very difficult to see how an isolated F3~V star could be responsible for \9173\ X-ray emission. Indeed, using the measurement given in \citet{2011Pavan} for the minimum 3~kpc distance derived in the present study, we find that the absorbed luminosity is about $6\times10^{30}$~\ergs, which is 2 to 3 orders of magnitude larger than the X-ray luminosity expected from the coronal activity of such a star \citep[see, e.g.,][]{1997Schmitt}, even without extinction correction. Alternatively, the star could be the companion of a compact object in a completely quiescent X-ray binary (XRB), which would reach such X-ray luminosity levels. For instance, the F6~IV/V star in GRO~J1655-40 was shown to fully dominate the system in quiescence \citep{1997Orosz, 2006Foellmi}. Nonetheless, quiescent BH-XRBs are usually found to have a relatively soft power law-like X-ray spectrum with $\Gamma\approx2$, without any cut-off, and quiescent NS-XRBs  have an X-ray spectrum dominated by the NS surface. It would moreover be an extraordinary coincidence to find a CV and an XRB within less than 3\arcsec\ from each other, perhaps located at the same distance. As such, we thus conclude that \9173\ is very likely to be a CV, contradicting its previous classification as an HMXB.

The optical spectrum of the most likely soft X-ray counterpart to \164\ is fully consistent with an isolated M2~III star. We carefully searched for emission lines that could prove the presence of an accretion stream but could not find any. This however does not rule out the possibility that it is the companion of a compact object. Indeed, it is very unlikely that an isolated M star would be detected by \intl. This statement is strengthened when considering that the 0.5--10~keV flux we derive, $2.0\times10^{-13}$~\ergcms, is equivalent to a luminosity of about $2.2\times10^{32}$~\ergs\ for a minimum 3~kpc distance, respectively, a level which would be abnormally high for coronal emission from an isolated low mass star \citep[see, e.g.,][]{2009Gudel}. Likewise, the relatively hard power law best-fit to the \chandra\ spectrum is not consistent with the expected soft X-ray emission from an isolated early M giant \citep[see, e.g.,][]{2014Rahouia, 2014Fornasini}. Instead, we rather believe that the most likely soft X-ray counterpart to \164\ is an X-ray faint symbiotic star, i.e. a white dwarf (WD) fed by the wind of its M2~III companion. The absence of shell-burning could lead to very weak optical emission lines undetectable through low-resolution spectroscopy, as recently proposed to explain the relatively bright hard X-ray emission of an M giant star thought to be isolated, SU Lyncis \citep{2016Mukai}. The authors argue that a potentially large population of such systems could be present in the Galaxy, leading to an underestimation of the known symbiotic stars. Although we stress that we may have picked the wrong soft X-ray counterpart, we believe that \164\ is a new member of this family.

\section{Conclusion}

We have reported on FORS2 spectroscopy of the optical counterparts of four IGR sources, the X-ray properties of which were consistent with what we expect from the elusive HMXBs that host BHs instead of NSs. We show that none of them is an actual HMXB, as \088\ and \9173\ are magnetic CVs, \381\ is a faint Seyfert 1.9 galaxy, and \164\ is, perhaps, a faint symbiotic star. On the one hand, our results illustrate the efficiency of ever-deeper X-ray surveys with soft and hard X-ray facilities such as \chandra\ and \intl\ in unveiling new populations of X-ray emitters with relatively hard X-ray spectra, in particular faint magnetic CVs and wind-accreting symbiotic stars. On a longer term, this will likely help us refine what we know about these sources and, perhaps, challenge current accepted schemes. However, our study also points towards the difficulty of identifying HMXBs , in particular SGXBs, and highlights the limitations of the accepted selection criteria of HMXB candidates -- a Galactic plane location, a hard X-ray spectrum, and a variable X-ray emission -- which must be revisited. It is thus likely that many hard X-ray sources thought to be HMXB candidates based on these criteria have another nature. If true, this should be taken into account in HMXBs and BH-BH population synthesis models. Likewise, the detection of new BH-HMXBs probably requires very sensitive hard X-ray instruments, as we expect them to be very weak accretors \citep[see, e.g.,][]{2004Zhang, 2014Casares, 2014Munar}. This is the reason why we favour the use of an hard X-ray observatory such as the {\it Nuclear Spectroscopic Telescope Array} \citep[{\it NuSTAR},][]{2013Harrison}, combined with state-of-the-art infrared spectroscopic facilities, to determine whether or not a significant population of BH-HMXBs exists in our Galaxy.

\section*{Acknowledgements}

We thank the referee for his/her comments and time committed to the correction of this work. FR thanks the ESO staff who performed the service observations. JAT acknowledges partial support from NASA under \xmm\ Guest Observer grant NNX15AW09G. RK acknowledges support from Russian Science Foundation through grant 14-22-00271. This research has made use of data obtained from the High Energy Astrophysics Science Archive Research Center
(HEASARC), provided by NASA's Goddard Space Flight Center. This research has made use of NASA's Astrophysics Data System, of the SIMBAD, and VizieR databases operated at CDS,
Strasbourg, France.




\bibliographystyle{mnras}
\bibliography{../mybib_tot}


\bsp	
\label{lastpage}

\end{document}